# One-dimensional model of streaking experiment in solids.

A.K.Kazansky[1,2], P.M.Echenique[2]

[1] *Fock Institute of Physics, The University of St.Petersburg, St.Petersburg, 198504, Russia*
[2] *Donostia International Physics Center (DIPC), Paseo Manuel de Lardizabal,
E-20018 San Sebastián/Donostia, Basque Country, Spain*
[3] *Departamento de Fisica de Maeriales and Centro Mixto CCSIG-UPV/EHU,
Facultad de Ciencias Quimicas, UPV/EHU Apdo 1072,
20080 San Sebastián/Donostia, Basque Country, Spain*

One-dimensional model for study of sub–femtosecond experiment with metal surface is put forward. The important features of the system, such as the pseudopotential for electron motion in the metal bulk, abrupt decrease of the normal to the surface external electromagnetic field in the bulk, finite value of the mean free path for electrons in the metal, and action on the ejected electron by the (stationary) screened positive hole in the metal are included in the model. The results obtained reveal dependence of the streaking effect on the final energy of the ejected electron. Meanwhile, the dependence of the streaking on the character of the initial state (localized or delocalized) appears to be more pronounced. This result may provide an additional mechanism for interpretation of the results of very recent experiment [Cavalieri *et.al*, NATURE **449** 1029-1032 2007 ].

Study of the real-time dynamics of electrons in condensed-matter systems is pertinent for progress in nanotechnology. The electron processes in nano systems are very fast and their investigation in real time requires application of experimental tools with sub–femtosecond time resolution. Recently the first experiment with streaking observation of electron dynamics in metal in the sub–femtosecond range was performed [1]. In this experiment, the surface of solid was illuminated by two pulses. The first pulse was a short XUV pulse with the frequency $\omega_X$ of about 90 eV and duration (FWHM for the field envelope) $\tau_X$ of about 0.2 fs. Intensity of this pulse is quite low. Another pulse was a relatively strong (power $W$ in $10^9$ - $10^{10}$ $W/cm^2$ range) near-infrared (NIR) laser pulse with the frequency $\omega_L$ about $1.5\,eV$, and with the duration $\tau_L$ about $10\,fs$. The energy spectra of the electrons ejected from the localized $f$-state and delocalized $d$-band through the (110) surface of tungsten in the direction normal to this surface were measured. The time delay between the two pulses was varied in [1] and the energy spectra of the ejected electrons were monitored as a function of this delay. These energy spectra are the result of steering of the electrons, ejected from the metal with the XUV pulse, by the electric field of the NIR pulse in vacuum. The energy acquired by the ejected electron from the NIR field depends on the time of the electron passage across the metal surface. Thus, measuring the dependence of the ejected electron energy spectrum on the time delay between the pulses, one can keep track on the process dynamics in the time domain. This is the idea of the 'streaking camera' [2]. Previously with this method a few experiments with isolated inert-gas atoms in the gas phase had been performed [3]. The streaking effect had been thus well established as an experimental tool and a theory for the experiments with isolated inert–gas atoms had been also developed [4]. The first experiment on a metal surface with time resolution in the attosecond range [1] has shown that the streaking method can be applied for such studies. This is a very important proof–of–principle experiment, in which a number of difficulties intrinsic for the streaking studies at the metal surfaces were overcome. Some preliminary theoretical studies [5,6] of non-stationary fast processes in metals have been also undertaken.

In [1], the experimental results were shown to be consistent with concepts derived from the static band structure. In the supplementary information to [1], the main approximations of the calculations were listed. A key issue remaining is the assumption that static band structures can be employed in this very dynamical process. Using a simple one dimensional model that includes selection rules and transition matrix elements and is fully time dependent we show that for these small times relevant to the experiment there is no time for the group velocity in the final state to be established and therefore the picture of static band structure is not valid. We also find that the delay time arising from the different character (localized versus delocalized) of the initial state is in agreement with the experimental results [1].

The processes triggered by instantaneous excitation of an electron in solid are very complicated and a number of various mechanisms can be of paramount importance. First, electrons in the metal are moving in the field of the lattice. This cfould in principle change the group velocity of the excited electron packet inside the bulk, as was described in [1]. Second, a localized electron after its ejection leaves in the bulk a positively charged hole which is then screened by the itinerant electrons. Third, the ejected electrons suffer inelastic collisions with electrons of the metal. This determines the depth from which the ejected electrons can reach the surface without inelastic collisions and thus carry a direct information on the processes in the bulk. Fourth, the normal component of the laser field decreases in the bulk abruptly to very small value under the condition of the experiment [1]. This determines the peculiarity of streaking effect in the system considered. In reference (1) the experimental results were shown to be consistent with concepts derived from the static band structure.

Here we formulate a simple and versatile model which includes these essential features of the phenomenon and allows one to estimate the magnitudes of the possible effects and analyze the influence of the parameters of the system on the output. The purpose of this study is to compute the delay time dependence of the energy spectra of electron ejected from a given initial state in the metal. Basically, two features are of importance in the theoretical investigation. First, the ejected electron moves in the lattice field and this could produce some difference in the ejected electron spectra [1]. Another point of attention is related to screening of the streaking IR field in the solid. Considering the first effect, one has to keep in mind that the electrons of few tens eV energy, being ejected from deep layers, suffer inelastic collisions with other electrons. The corresponding mean free path (MFP) is about 5-7 Å [7]. Thus, experimentally information can be obtained only on the events occurring in a very few external layers of the metal. On the other hand, variation in streaking for the electrons ejected from various layers can be observed only if the normal component of the streaking field inside the bulk is promptly screened to quite small value and thus is not uniform. The frequency of NIR laser, $1.5\,eV$, is lower than the plasmon frequency in metals and thus the screening of the field normal to the surface is similar to screening of static electric field, that is the penetration depth is in the 2-3 Å range [8]. The interplay between the MFP and screening length restricts substantially the experimental window for observation of peculiarities of attosecond streaking in metals: the screening length must be less than the MFP.

In our model, we consider a 1D slab of the metal of thickness $L = 300\,a.u.$ The time evolution of the ejected electron wave packet $\Psi(z,t)$ is governed by the non-stationary Schrödinger equation (atomic units are used)

$$\imath\frac{\partial}{\partial t}\Psi(z,t) = \Big( -\frac{1}{2}\frac{\partial^2}{\partial z^2} + (U_s(z) - E) + U_h(z,t) -$$
$$\imath\,\gamma(z)\,\Big)\Psi(z,t) + E_L(z)\,\epsilon_L(t)\cos(\omega_L t + \phi)\Psi(z,t) + \quad(1)$$
$$\hat{V}_X(z,t)\,\Phi_0(z), \quad z > -L.$$

This equation is written within the rotating wave approximation (RWA) for the weak XUV pulse (for some details, see [4]). The quantity $E$ is the energy (with respect to the vacuum level) at the center of the energy spectrum of the ejected electron. It is a sum of the electron energy in the initial state, $-|E_0|$, and the carrier frequency $\omega_X$ of the XUV pulse. The wave function of the electron in the initial state is $\Phi_0(z)$. The term $\hat{V}_X(z,t)\,\Phi_0(z)$ in Eq.(1) describes interaction of the XUV pulse with the initial electron state; it is taken in the length form $\hat{V}_X(z,t)\,\Phi_0(z) = \epsilon_X(t)\,z\,\Phi_0(z)$. Here $\epsilon_X(t)$ is an envelope of the electric field of the XUV pulse. We assume it to be Gaussian $\epsilon_X(t) = \exp(-(t - t_{delay})^2/\overline{\tau}_X^2)$ with $\overline{\tau}_X = 125\,a.u.$, FWHM $= \tau_X = 0.21\,fs$. The $E_L(z)$ in Eq.(1) describes the NIR laser electric field:

$$E_L(z) = \begin{cases} \xi + (z - z_{im}) & z > z_{im} \\ \xi\exp((z - z_{im})/\xi) & z < z_{im} \end{cases} \quad(2)$$

with the screening length $\xi = 4\,a.u.$ [6]. The parameter $z_{im}$ in Eq.(4) is a position of the image plane, it enters the parametrization of the pseudopotential $U_s(z)$ (see [5], Eq.(2)-(5)). In the present case $z_{im} = 2.105\,a.u.$ In the experiment [1] the streaking field was incident on the surface at the Brewster's angle ($\sim 75°$). Thus, the normal component of this field in the metal is about 16 times weaker than the incident field. In our model we set the streaking field in the metal equal to zero.

The envelope of the NIR pulse is taken as $\epsilon_L(t) = 0.5\mathcal{E}_0\{1 - \cos[\pi t/\tau_L]\}$ with the FWHM $\tau_L = 5\,fs$ and $\omega_L = 1.6\,eV$, $\phi = 0$ in Eq.(1). The field strength $\mathcal{E}_0$ is conventionally related to the intensity of the pulse $W$. It was quite difficult to control the intensity of the NIR laser pulse in the experiment [1]. In the previous experiments in the gas phase, the intensity $W$ was greater than $10^{13}\,W/cm^2$. In experiments with the solids the intensity is restricted by a number of requirements to be much smaller than in the gas phase. By comparing the experimental results with the results of our computation, we have estimated the intensity of the NIR laser field in the experiment [1] as $W = 6\,10^9\,W/cm^2$.

The (pseudo)potential $U_s(z)$ in Eq.(1) mocks the interaction of an electron with the lattice in a finite metal. For this potential, we take the parametrization by Chulkov et al [9] with the parameters corresponding to the Cu(111) surface. The function $U_h(z,t)$ in Eq.(1) describes the interaction between the ejected electron and the hole left in the solid by this electron. We assume a static approximation for this potential as a screened softened Yukawa potential with the screening length $\xi$, the same is in Eq.(2):

$$U_h(z) = \exp(-|z - z_{at}|/\xi)/\sqrt{(z - z_{at})^2 + a_0^2} \quad(3)$$

with $a_0 = 0.4\,a.u.$ Softening of this interaction allows us to apply in solving the non-stationary task more advantageous Fast Fourier algorithm. Our study can adopt any given time-dependent screening of the hole. The damping function $\gamma(z)$ in Eq.(1) is non-zero at $z < z_{im}$, where $\gamma(z)$ for the electron with energy $E$ in the bulk is approximated as $\sqrt{2E}/2\lambda_f$ with $\lambda_f$ being electron elastic MFP [7] near the metal surfaces. The quantity $\lambda_f$ effectively takes into account attenuation of the ejected electron wave packet by inelastic collisions with other electrons in the bulk. It weakly depends on the electron energy, being close to 5 Å at a few tens eV range [8]. We have set $\lambda_f = 10\, a.u.$. The initial wave function $\Phi_0(z)$ for the case of initial localized state has been computed as an eigenfunction of the potential $U_s(z) + U_h(z)$ localized in a vicinity of an atom placed at $z_{at}$ (the atoms are nested at the maxima of the potential energy $U_s(z)$). For ionization from an initial delocalized state, the wave function $\Phi_0(z)$ has been computed as an eigenfunction of the potential $U_s(z)$, in this case the term $U_h(z)$ in Eq.(1) has been omitted in the wave propagation as well. Note that the initial electron states are assumed to be below Fermi level. They are not perturbed by inelastic collisions with other electrons and in the computation of the initial functions $\Phi_0(z)$ the damping $\gamma(z)$ must be omitted. In Fig.1 the relevant functions of the model are plotted.

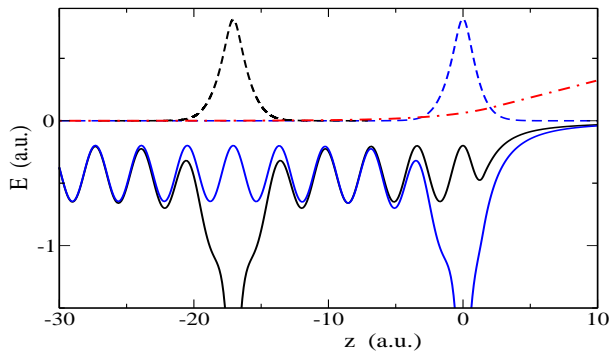

FIG. 1: The potential energies and the wave functions of the localised states. Solid lines: the potentials in the bulk $U_s(z)$ with account for attraction of the electron to the positive hole, for electrons ejected from the topmost - blue line, for electron ejected from the 5th atom - black line; the dashed lines show the corresponding initial wave functions $\Phi_0(z)$; dash-dot (red) line shows the profile of the laser pulse potential $E_L(z)$ screened in the bulk.

Our computations have been performed with the split-propagation algorithm with the fast-Fourier computation of the kinetic energy. The mesh comprises 16384 knots and covers uniformly the interval $z \in [-300\, a.u., 1340\, a.u.]$. The time step has been set equal to $0.03\, a.u.$. The outgoing wave asymptotic condition has been provided by the artificial adsorbing potentials at the edges of the mesh. Propagation has been performed till $t_{fin} = \tau_L + 10\, a.u.$, when the wave packet has left the region where the potential from the metal is noticeable, but the principle part of the wave packet is yet pretty far from the region where the absorbing potential starts to work. The amplitudes of the final states population have been computed with the Fourier transform of the final wave packet. It is important that, due to the inelastic collisions in the bulk, a part of the wave packet there is negligibly small at $t = t_{fin}$.

Let us first consider excitation of the electrons from localized initial states. This case corresponds to the ejec-tion of electrons from the 4f band in the experiment [1]. In this case, the energy spectrum is computed as a sum of the spectra of the electrons ejected from the atoms at various positions in the bulk, we take into account 17 consecutive atoms. Due to the inelastic collisions, the relevant electron yield decreases exponentially with the depth of the atom, following the dependence $I = I_0 \exp(-|z_{at}|/\lambda_f)$, where $I_0$ is the total electron yield from the topmost atom. The ejected electron spectra from various atoms demonstrate general behavior: they are almost Gaussian and each of them can be described with the position of the 'center-of-mass' (COM) of the spectra [1], their FWHM, and the integrated yield. The first two features demonstrated the universal for the streaking effect behaviour [2]. The dependence of COM on the delay time follows the time-dependence of the vector-potential of the NIR field [2]. The FWHM of the spectra oscillates in phase with the applied electric field [10]. As an specific effect due to the screening of the NIR field in the solid, the spectra from various atoms are slightly shifted with respect to each other, since and it is the time when the electron escapes the bulk determines the process.

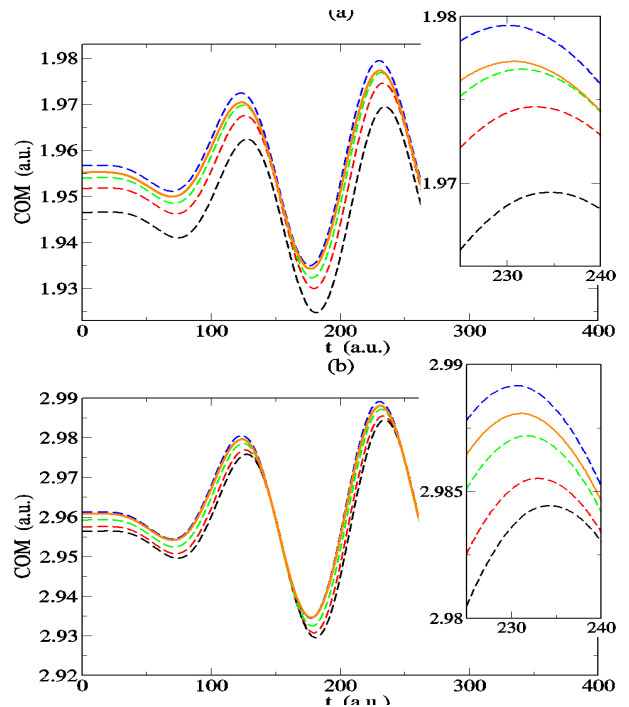

FIG. 2: The 'center-of-mass' (COM) of the spectra of ejected electrons; a: $E = 2$ a.u., b: $E = 3$ a.u. Black dashed line: the topmost atom, red dashed line: the 2nd atom, green dashed line: the 3rd atom, blue dashed line: the 4th atom. The orange solid line: the COM of the total spectrum.

In order to isolate the effect of the final state energy for a fixed initial state, in Fig.2, the dependence of COM on the delay time for the two considered final energies are shown. The curves for the atom at the very sur-face are well synchronized. Then, the curves for deeper atoms are shifted to the earlier times. The shift depends both on the position of the atom and on the energy $E$ considered. The reason of this shift is obvious: the electron ejected from an atom in the solid at time $t$ starts to feel the laser electric field only at $t' = t + \tau_c(z_{at}; E)$, where $\tau_c$ is the time for electron passing from $z_{at}$ to the surface $z = 0$. Noteworthy that the curves in Fig.2 for $E = 3$ a.u. are more dense than for $E = 2$ a.u., this means that the electrons with larger energy move faster in the bulk. The position of the COM for the sum of the spectra of electrons ejected from different atoms corresponds roughly to the COM of the spectrum ejected from the third atom, this atom is placed at the distance close to $\lambda_f$ from the surface. It is then obvious that the COM for the higher final energy is shifted to the right with respect to the COM of the spectrum for lower energy. The shift in our case is about $12\,as$. The shifts of the extrema of the COM curves for different atoms with respect to the COM from the topmost atom can be effectively represented as $\tau_c(z_{at}) = z_{at}/v(E)$. The effective velocities $v(E)$ read: $v(E = 3\,a.u.) = 2.68 \pm 0.01\,a.u.$ and $v(E = 2\,a.u.) = 2.29 \pm 0.01\,a.u.$. These values of the velocities give for the effective potential energy in the bulk $U_{av} = E - v^2(E)/2$ the values - 0.62 $a.u.$ for $E = 2\,a.u.$ and - 0.60 $a.u.$ for $E = 3\,a.u.$ These values are quite close to each other and are in good correspondence with the minimal value of the potential energy $U_s(z)$ in the bulk. The remaining small deviation in the effective potential energies for different final energies is related to a weak effect of attraction between the positive hole and the ejected electron, which slightly depends on the electron energy, giving a deeper effective potential for a slower electron. These results evidence that the effect of the group velocity does not reveal itself within the present model. Actually, the group velocity in the bulk is formed by the interference of the wavelets arising from scattering of the incident wave by the atoms in the lattice. Taking into account a small value of the MFP in the metal at the final energies considered, one may conclude that under these conditions the group velocity of the wave packet in the bulk hardly can be formed. In our computations, the forming of the group velocity could reveal itself in a non-uniform dependence of the time $\tau_c(z_{at}; E)$ on the atom distance from the surface. We do not notice such an effect in our data.

Now let us turn to the problem of ionization from the continuous band. This band for the considered system covers the interval from $-12.27\,eV$ to $-5.92\,eV$. in a restricted metal this continuum is represented with 75 discrete states. As a center of the continuous band we take the 45th state. This state has an energy $\epsilon_{45} = -9.46\,eV$. We took as initial 15 states in the band and propagated them in time with Eq.(1) taking various energies in vacuum $E(k) = E_0 + \epsilon_n - \epsilon_{45}$ for the various initial states $\Phi_k(z)$. This corresponds to the excitation of the band by such XUV pulse that the center of the spectrum of electrons ejected from the center of the allowed band lies at 3 $a.u.$ with respect to the vacuum level. First, the electron spectrum obtained with excitation from the selected states from the band have been calculated. Then the total spectrum was obtained with the integration over the entire energy band of the delocalized electrons. The results of computations of the reduced center-of-mass of the total ejected electron spectrum in dependence on the delay between XUV and the laser pulses are shown in Fig.3 for a few initial states. One can see that the results are practically unshifted to each other.

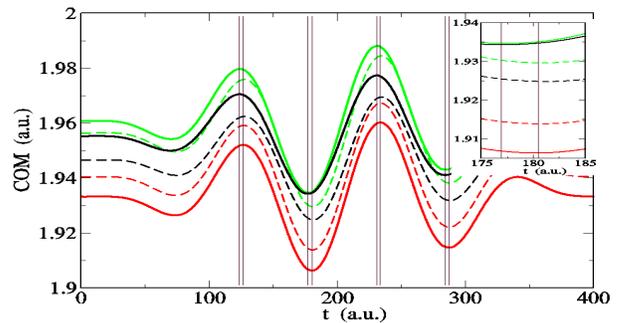

FIG. 3: The shifted to the same scale the COMs for three various cases. Solid lines: COMs of the total spectra; red: initial delocalized states; black: localized state, $E = 2\,a.u.$; green: localized state, $E = 3\,a.u.$. Dashed lines: black and green - contributions from the topmost atom, red: COM for the highest state of the band.

The principle results obtained within the present study are collected in Fig. 3. One may see that the COMs from the first atom are synchronized pretty well. The COMs obtained with the initial states from the band are almost synchronized with the COMs obtained from the topmost atom. The COMs for the spectra calculated with the initial localized states are noticeably delayed with respect to the COMs of spectra for the delocalized states. This shift is about 85 as, while the shift obtained here for the COM from localized states at final energies E=2 a.u. and E=3 a.u., that is about 10 as. The experimental result [1] is $110 \pm 70$ as.

In conclusion, the results of a time-dependent approach that goes beyond the static model [1] are presented. Our model includes the main ingredients of the short time physics involved in the experiment [1]. Although it is a one dimensional model and uses an energy independent pseudopotential that does not take into account the fact that the electrons in the final state would penetrate different regions of the ionic cores and therefore feel an energy dependent interaction potential. In addition, our model includes approximately the time required to fill the hole but neglects the hole in the delocalized state. These effects should be investigated further. What would be needed is a full time-dependent version of the calcula-tions [1] with explicit inclusion of the time required to fill the hole. This is not computationally feasible at the moment. However the calculations presented here, despite using a one-dimensional model potential, use a time dependent approach and take into account all relevant effects leading to a reasonable agreement with experiment. They also point out, in agreement with what the authors [1] state as a weakness of their model in the Supplementary Information to this reference, that the assumption that static band structure can be used in these very short times is not valid.

AKK deeply acknowledges the Ikerbasque Fellowship. PEM acknowledges partial support from the University of the Basque Country (9/UPV 00206.215-13639/2001), the Basque Unibersitate eta Ikerketa Saila and the Spanish Ministerio de Education y Ciencia (MEC) (FIS 2004-06490-C03-01 and CSD2006-53). We are thankful to U Heinzmann and N Muller for useful discussions and to M I Stockman and N M Kabachnik for critical comments.